\documentclass[10pt,a4paper,twocolumn]{article}
\usepackage{lineno}
 \usepackage{nopageno}

\usepackage{lipsum}
\usepackage{dblfloatfix}
\usepackage{authblk}
\usepackage{graphicx}

\usepackage{svg}
\usepackage[explicit]{titlesec}
\usepackage[labelfont=bf,labelsep=endash,font=footnotesize]{caption}
\usepackage{tabu}
\usepackage{xcolor}
\usepackage[
    backend=biber, 
    natbib=true,
    style=numeric,
    maxnames=50,
    sorting=none
]{biblatex}
\usepackage[hidelinks]{hyperref}
\usepackage[hang]{footmisc}
\usepackage[normalem]{ulem}
\usepackage[top=2.5cm, bottom=2.8cm, left=1.5cm, right=1.5cm]{geometry}
\usepackage{abstract}
\usepackage{mathtools}
\usepackage{balance}
\usepackage{wrapfig}
\usepackage{tabularx}
\usepackage{algorithm}
\usepackage{algpseudocode}
\usepackage{bm}

\usepackage{amsmath}
\usepackage{amsfonts}
\usepackage{hyperref}
\usepackage{graphicx}
\usepackage{svg}
\usepackage{pifont}
\usepackage{todonotes}

\usepackage{subcaption}
\newcommand{\cmark}{\ding{51}}%
\newcommand{\xmark}{\ding{55}}%

\newcommand{\algrule}[1][.2pt]{\par\vskip.5\baselineskip\hrule height #1\par\vskip.5\baselineskip}
\makeatletter
\renewcommand\AB@affilsepx{, \protect\Affilfont}
\makeatother

\providecommand{\keywords}[1]{\textbf{Keywords}\ \ \textendash\ \   #1}

\titleformat{\section}{\large\bfseries}{\thesection.}{1em}{\MakeUppercase{#1}}
\titlespacing*{\section}{0pt}{12pt}{6pt}

\titleformat{\subsection}{\large}{\thesubsection}{1em}{#1}
\titlespacing*{\subsection}{0pt}{12pt}{6pt}

\titleformat{\subsubsection}{\large\itshape}{\thesubsubsection}{1em}{#1}
\titlespacing*{\subsubsection}{0pt}{12pt}{6pt}

\newcommand{\ITUurl}[1]{\textcolor{blue}{\urlstyle{same}\url{#1}}}

\newcommand{\ITUpar}{\vspace{8pt}\par}

\newcommand{\ITUfootnote}[1]{\footnote{#1}}

\renewenvironment{abstract}
               {\list{}{
               \setlength{\rightmargin}{0mm}
               \setlength{\leftmargin}{0mm}
               \vspace{-0.25in}
                \item[\textit{\textbf{\hspace{22pt}Abstract  }}  \textendash]\relax}}
               {\endlist}

\def\starttable{\vspace{6pt}\begin{table}[ht]\center}
\def\startfigure{\vspace{6pt}\begin{figure}[ht]\center}

\makeatletter
\def\tagform@#1{\maketag@@@{\ignorespaces#1\unskip\@@italiccorr}}
\makeatother

\title{\large{\MakeUppercase{\textbf{QT-Routenet: Improved GNN Generalization to Larger 5G Networks By Fine-Tuning Predictions From Queueing Theory}}}}

\author[1]{\normalsize{Bruno Klaus de Aquino Afonso}}
\author[1]{\normalsize{Lilian Berton}}

\affil[1]{\normalsize{Federal University of São Paulo (ICT-UNIFESP)}}
\date{\vspace{-12pt}{\small NOTE: Originally published in \url{https://www.itu.int/pub/S-JNL-VOL3.ISSUE3-2022-A05} \\Corresponding author: Bruno Klaus de Aquino Afonso, bruno.klaus@unifesp.br} \\  \endgraf\rule{\textwidth}{1pt}}

\begin{document}
%% this should be commented out if line numbers in the text are not wanted

%%= title and abstract =%%
\twocolumn[

\begin{@twocolumnfalse}
\maketitle

%%= abstract text � SUBSTITUTE HERE! =%%
\begin{abstract}
\textit{
%This template provides detailed instructions for submitting papers to ITU J-FET. Papers must be in English and in black font, no page limit. The abstract should appear in italics at the top, below the title and author's area.  The abstract should normally contain 150 to 200 words, and in no case shall it exceed 300 words. All abbreviations and acronyms used in the abstract should be defined, and in the text the first time used. Do not cite references in the abstract.
%%138 words:
In order to promote the use of machine learning in 5G, the International Telecommunication Union (ITU) proposed in 2021 the second edition of the ITU AI/ML in 5G challenge, with over 1600 participants from 82 countries. This work details the second place solution overall, which is also the winning solution of the Graph Neural Networking Challenge 2021.  We tackle the problem of generalization when applying a model to a 5G network that may have longer paths and larger link capacities than the ones observed in training. To achieve this, we propose to first extract robust features related to Queueing Theory (QT), and then fine-tune the analytical baseline prediction using a modification of the Routenet Graph Neural Network (GNN) model. The proposed solution generalizes much better than simply using Routenet, and manages to reduce the analytical baseline’s 10.42 mean absolute percent error to 1.45 (1.27 with an ensemble). This suggests that making small changes to an approximate model that is known to be robust can be an effective way to improve accuracy without compromising generalization.
}
\end{abstract}
%%======%%

\ITUpar
%%= keywords =%%
\keywords{5G networks, fine-tuning, graph neural network, ITU challenge, queueing theory}
%%======%%

%\ITUnote{Title, abstract and keywords must be identical to the ones submitted electronically in EDAS \textendash\ Editor's Assistant. Use the command \texttt{\textbackslash ITUnote} to achieve the appropriate formatting.}
\ITUpar

\end{@twocolumnfalse}
]

%%= start of sectioning � MODIFY EACH TITLE AND LABEL AS YOU PLEASE =%%
\section{Introduction} 
\label{sec:intro}
During the year of 2021, the International Telecommunication Union (ITU) once again brought to the forefront the use of machine learning as a means to maximize the efficiency of 5G. The second edition of the \textit{``ITU AI/ML in 5G challenge''}  introduced a diverse set of challenges related to the development and training of machine learning models to solve particular problems within the realm of 5G networks. Over 1600 competitors from 82 countries were asked to solve problems that were put forth by hosts from different regions \cite{ITUnumbers}. This work details the first place solution of the challenge proposed by the Barcelona Neural Networking Center, named \textit{Graph Neural Networking Challenge 2021 - Creating a Scalable Digital Network Twin}, a.k.a. \textbf{GNNet Challenge 2021}. This solution would later on compete against winning solutions from other regional hosts in the Grand Challenge Finale, ending up with second place overall.

\ITUpar Much like in the previous year, the GNNet challenge was centered around creating a predictive model for 5G networks: given a topology and routing configuration, one must predict the per-path-delay. In addition, the GNNet Challenge 2021 had a specific goal in mind: to address the current limitations of Graph Neural Network (GNN) architectures, whose generalization suffers greatly when predicting on larger graphs. This was verified by the organizers to be the case for \textit{RouteNet} \cite{rusek2019unveiling}, a message-passing GNN model  that influenced most solutions from the 2020 edition of the challenge \cite{gnnet2020}. When creating a dataset, it is usually not feasible to gather data from a currently deployed network, as that would require us to explore edge cases that directly lead to service disruption, such as link failures. The alternative is to generate everything from a small network testbed created in the vendor's lab. The distribution of the graphs observed in validation/test set are therefore different from the ones observed in training.
\begin{table}[!h]
\caption{Types of approaches}
\scriptsize
\begin{tabularx}{0.45\textwidth}{l|X|X|X}
                                 & \textbf{Fast Enough?} & \textbf{Top tier results on small graph?} & \textbf{Generalizes to larger graphs?} \\ \hline
\textbf{Analytical}   & \cmark                               & \xmark                                 & \cmark                                    \\
\textbf{Packet simulators} & \xmark                  & \cmark                                & \cmark                                    \\
\textbf{RouteNet}                & \cmark                               & \cmark                                & \xmark                                     \\
\textbf{Proposed solution}       & \cmark                               & \cmark                                & \cmark                                   
\end{tabularx}
\label{tbl:1}
\end{table}
\ITUpar To understand the solution detailed in this report, it is helpful to look at previous approaches (Table \ref{tbl:1}). They are divided into 3 categories:  analytical approximations, packet simulators and RouteNet, a model based on message-passing GNNs. Packet simulators were not allowed in the competition in principle due to excessive running times; analytical approaches generalize well and run fast, but they do not offer competitive performance;  RouteNet is still fast and more accurate than analytical approaches, but fails to generalize to larger graphs. For our proposed solution, we \textbf{extract invariant features from the analytical approach}, and feed them to a GNN. This way, we can \textbf{maintain generalization while outperforming the purely analytical} approach. \ITUpar

\section{RELATED WORK}
\label{sec:related_work}
The problem of predicting traffic in networks has been long studied within Queueing Theory (QT) \cite{zukerman2013introduction}, a branch of mathematics that deals with the analysis of waiting lines. In a queueing system, customers randomly arrive at a certain place to receive a service, and then leave upon its completion. By modeling the arrival process and service of customers with probability distributions, we can use QT to estimate Key Performance Indicators (KPIs) such as delay and jitter.
\ITUpar
Within the context of 5G networks, we are interested in modeling the arrival and service of network packets. We must model each link of the network as a separate queue. 
The simplest case is the M/M/1 queue, where the arrival process is Poisson, the service process is exponential, and there is one server. More intricate models include the M/M/1/B queue, which has a buffer that can hold up to B items. 

\ITUpar Traffic flow will be heavily dependent on the routing algorithm and network topology. Given a model of each link as a queue, one can derive a system of equations related to traffic balance on the network. By solving those equations, one arrives at the analytical solution for the relevant performance metrics. This provides us a way to analyze these systems with a solid theoretical foundation. However, analytic models used to predict KPIs in large-scale networks often make unrealistic assumptions about the network, and as a result are not accurate enough.

\ITUpar  If the time required to compute the KPI estimate is not a concern, it is interesting to consider packet simulators such as OMNeT++\cite{omnet}.  A model in OMNeT++ consists of nested modules that communicate by message passing. The topology of the model is specified by a topology description language. In this setting, analytical intractability is not a concern, and one can get more accurate results by simulating individual packets. However, this comes at a high cost when you consider the running time. According to the organizers of the  GNNet Challenge 2021, packet simulators were used to help create the competition's dataset. Combined with their excessive running times, it is no surprise that they were prohibited.

\ITUpar Machine learning is a powerful tool that can help us achieve better results than we would get through analytical methods alone. By using deep neural networks, we can learn the intricacies of real-world networks by leveraging huge amounts of data. As the input of our model is a network, the problem is very suitable to graph neural networks \cite{gnnsurvey}.
\ITUpar Routenet \cite{rusek2019unveiling} is the machine learning approach most influential to our work. It is a GNN-based model that uses update functions to maintain and update representations during message-passing iterations between links and paths. In addition to working with a natural network representation, Routenet is able to relate topology, routing, and input traffic in order to accurately estimate KPIs. Routenet outperforms the analytical baseline even when the latter was particularly suited to the dataset. In addition, it can handle different network topologies than the ones observed in training. 

\ITUpar After the initial promising results, more experimentation was conducted in \cite{happexploring} to evaluate the generalization capabilities of Routenet. The authors found that, when the evaluation data is drastically different from the training data, Routenet's predictions get significantly worse. This includes, but is not limited to, larger link capacities and longer paths.

\ITUpar During the Graph Neural Networking Challenge 2021, many approaches were devised to solve this generalization problem. At the time of writing, one available example is the data augmentation of \cite{ferriol2021scaling}, where link capacities are defined as a product of a virtual reference link capacity and a scaling factor. 

\section{FRAMEWORK}
\label{sec:sec3}
Our model was built from scratch in the Python language using Pytorch 1.8.1 \cite{pytorch} and Pytorch Geometric 1.7.0 \cite{pytorch_geometric}. The code requires a GPU; we used an RTX 3080 with 10GB VRAM. Moreover, 16GB of RAM is enough to not run out of memory. Our code was divided into 3 different Jupyter notebooks: one for creating the dataset, two for \textbf{2 similar models} whose \textbf{average }constituted the final prediction used for this challenge. The source code and frozen model weights are available on  Github \ITUfootnote{\small \url{https://github.com/ITU-AI-ML-in-5G-Challenge/ITU-ML5G-PS-001-PARANA}}.
\subsection{Input format}

\begin{table}[!ht]
\caption{Converted dataset information}
\begin{tabular}{l|l|l}
                      & \textbf{Samples} & \textbf{Network size}\\ \hline
\textbf{Training}   & 120000 & 25-50 nodes                   \\
\textbf{Validation} &3120    & 51-300 nodes                  \\
\textbf{Test}       & 1560    & 51-300 nodes                 
\end{tabular}
\label{tbl:2}
\end{table}
Because we are interested in generalization to larger networks, the samples in the validation dataset  are considerably larger. As shown in Table \ref{tbl:2}, the networks seen in training have at most 50 nodes, whereas those in the validation set may have up to 300.

\ITUpar The validation set is divided into 3 subsets. All subsets were provided to participants by the competition organizers, and each one captures a type of network that differs from the training set. In Subset 1, longer paths are artificially generated, and only those paths transmit traffic. Subset 2 uses variants of a shortest path routing policy, with all source-destination paths producing traffic. To make up for the increased traffic, the routing  includes links with larger capacity than the ones encountered in the training data.  Lastly, Subset 3 can be considered a combination of the previous two: it uses routing schemes with larger paths, and also larger link capacities. The test set is assumed to have the same distribution as the validation set.

\ITUpar In order to improve speed when training our model, we modify the script provided to challenge participants\ITUfootnote{\url{https://github.com/BNN-UPC/GNNetworkingChallenge/tree/2021\_Routenet\_TF}}. Our script allows one to run multiple processes in parallel, to speed up the creation of this ``converted'' dataset. Due to the large amount of files, we recommend an SSD with at least 60GB of memory available. Each flow between two nodes is considered as a separate entity, with the following attributes: 
\begin{enumerate}
\itemsep0em
    \item \texttt{p\_AvgPktsLambda}: Average number of packets of average size generated per time unit.
    \item \texttt{p\_EqLambda} Average bit rate per time unit.
    \item \texttt{p\_AvgBw}: Average bandwidth between nodes (bits/time unit).
    \item \texttt{p\_PktsGen}: Packets generated between nodes (packets/time unit).
    \item \texttt{p\_TotalPktsGen}: Total number of packets generated during the simulation.
\end{enumerate}

We zero out \texttt{p\_TotalPktsGen}, as we did not find the inclusion of simulation time  helpful. In order to subject the network to \textbf{more varied input values}, we divide link capacity \texttt{l\_LinkCapacity} (the sole link attribute) and all path attributes by \texttt{p\_AvgPktsLambda} before normalization. 
\begin{figure}[!ht]
\centering
\includegraphics[width=0.2\textwidth]{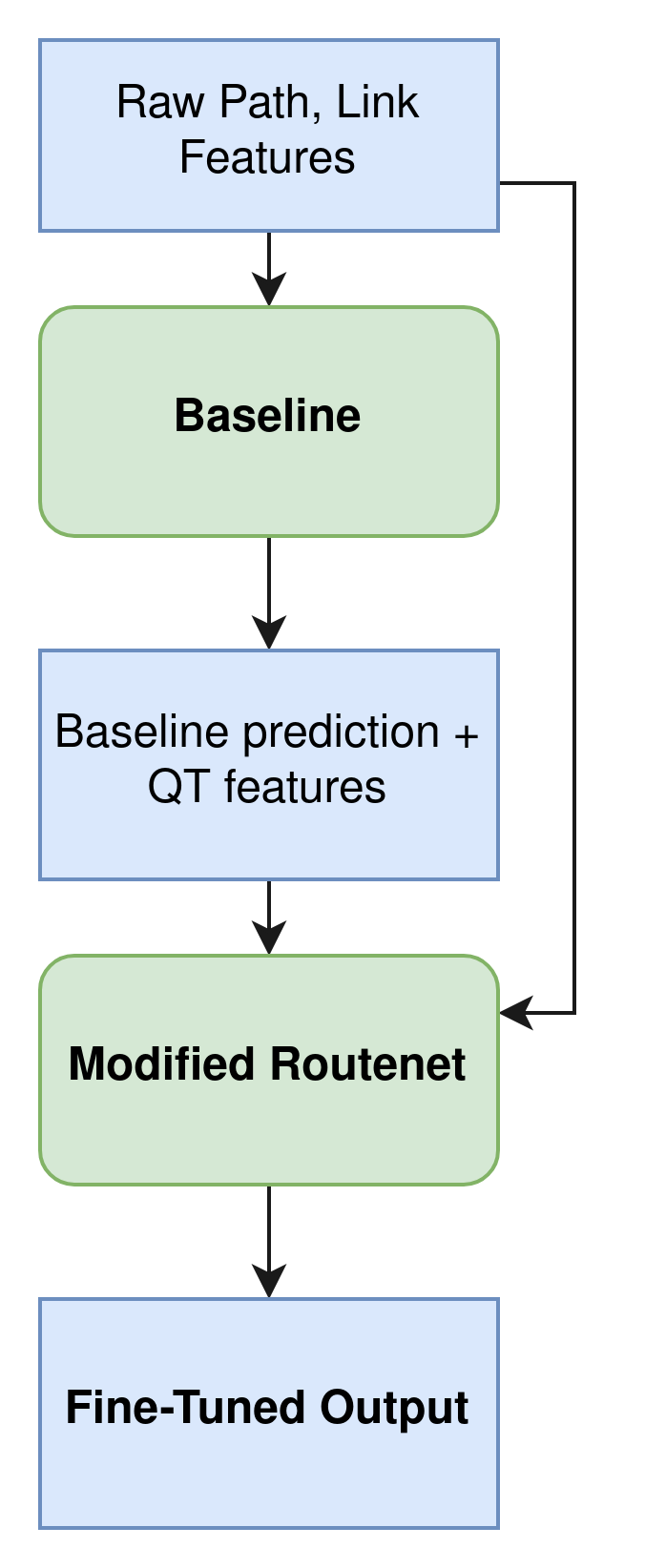}
\caption{Basic steps taken in our model. Instead of directly using the original ``raw'' path and link features, we feed them to the QT baseline and extract higher-level  features, including a reliable approximate prediction. A modified Routenet fine-tunes this prediction to improve the QT baseline while still maintaining generalization.}
\label{fig:model}
\end{figure}
\subsection{QT-Routenet: Model overview}

\ITUpar
The idea behind the proposed QT-Routenet model is very simple in principle, combining Routenet and smart feature extraction. In Fig. \ref{fig:model}, we can see that the original features are fed to the baseline, resulting in higher-level QT features, e.g. the baseline prediction. The higher-level features (and, optionally, the original features) are then fed to a modified Routenet to obtain  fine-tuned output. 

\ITUpar 
By experimentation, it was verified by the organizers that simply feeding the raw path and link features to the existing Routenet architecture leads to over $300\%$ \textbf{Mean Absolute Percent Error (MAPE)} in the competition dataset \cite{rnError}. This is likely because the distribution of those features in the validation and test sets is much different than the one observed in training. On the other hand, the baseline prediction remains reasonably consistent between the training set and all validation sets. This motivated us to tackle the problem at the input level: using the baseline prediction (and other QT features), we wanted to stay as conservative and close to the baseline as possible, but still use a graph neural network to obtain improved results.
\ITUpar One could potentially describe our approach as using queueing theory to assign an initial (imperfect) label, and then learning to smooth it adequately over the network. However, we should always refer to the baseline prediction as a feature, not a label. The reason is twofold: first, the actual label in this problem is the path delay obtained in the simulation conducted by the organizers; secondly, the prediction will be available for all future data, which avoids some overfitting issues when using known labels to optimize graph models with gradient descent \cite{aquino2021optimizing}.

\subsection{Defining the heterogeneous graph}
Our GNN model works with a matrix $\bm{X}$, which is the concatenation of all attributes. The number of rows is equal to the sum of the number of paths, links and nodes. These columns all remain fixed, so we also need to add ``hidden state'' columns to each of these entities. These columns are used to perform message-passing, taking in information about other ``hidden state'' columns and also fixed columns. We denote by $\bm{X}_P,\bm{X}_L,\bm{X}_N$ the fixed columns of paths, links, and nodes.  The hidden state columns are zero-initialized and denoted by  $\bm{X}_{Ph},\bm{X}_{Lh},\bm{X}_{Nh}$. We intended to put some provided global attributes into $\bm{X}_N$ but eventually opted for setting them to zero. We standardize all features before feeding them to the model.
\ITUpar

Next, we must go over network topology. We denote the topology by $\bm{E}$. The conditions for the existence of edges are:
\begin{itemize}
\itemsep0em
    \item $\bm{E}_{PL}$: Whenever a link is part of a path
    \item $\bm{E}_{PN}$: Whenever a node is part of a path
    \item $\bm{E}_{LN}$: Whenever a node is part of a link
\end{itemize}
In practice, edges are directed. We may use the  terminology  $\bm{E}_{PL}$ to indicate that the direction is \textit{path-to-link}, whereas $\bm{E}_{LP}$ is \textit{link-to-path}. In addition, our code contains a special function, \texttt{SeparateEdgeTimeSteps}, which is able to output a list separating $\bm{E}_{LP}$. The $k$-th element of this list has all of the $\bm{E}_{LP}$ edges satisfying a condition: that the link is the $k$-th one found while traversing the path. This separation allows us to preserve order information and use recurrent layers. 

\subsection{Message passing model}
There are two message passing models, which are listed as Algorithm \ref{alg:model_1} and Algorithm \ref{alg:model_2} \textbf{(see Appendix)}. The main difference is that the first model includes nodes in the message mechanism, which are ignored by the second model. For both models, we feed the initial input matrix $\bm{X}$ to a \textbf{Multilayer Perceptron (MLP)}. Then, we perform a number of message-passing iterations using convolutions. First the path entities receive messages, then nodes, and lastly links.
\ITUpar Messages are exchanged from links to paths using a single \textbf{Chebyshev Graph Convolutional Gated Recurrent Unit Cell} \cite{seo2018structured} layer imported from {Pytorch Geometric Temporal} \cite{rozemberczki2021pytorch}. All other convolutions were set to be \textbf{Graph Attention (GAT)} \cite{velivckovic2017graph} layers, and different GAT layers are used for each iteration. We set the first few hidden columns to be equal to the baseline features, so that this information is preserved similarly to the other fixed features. After the message-passing rounds, we feed $\bm{X}_{L}$ and  $\bm{X}_{Lh}$ to another MLP to obtain the prediction for \textit{average queue utilization}.  Finally, the average path delay is obtained using the formula
\begin{equation}
    \texttt{pathDelay} \approx \sum_{i=0}^{\texttt{n{\textunderscore}links} } \texttt{delayLink}(i)
    \label{eqn:path_delay}
\end{equation}
The delay on each link includes both the time waiting in the queue, as well as the time actually passing through the link. The former is given as
\begin{equation}
 \texttt{queue\_delay}_{i} = \frac{\texttt{avg\_utilization}_{i} \times \texttt{queue\_size}_{i}}{\texttt{link\_capacity}_{i}}\end{equation}
whereas the latter can be approximated as:
\begin{equation}
    \texttt{transmission\_delay}_{i} = \frac{\texttt{mean\_packet\_size}}{\texttt{link\_capacity}_i}
\end{equation}
From there, we calculate the average utilization of the link:
\begin{equation}
    \texttt{avg\_utilization}_i = \sum_{j=0}^{b_i} j(\pi_{0}\rho_i^j)
\end{equation}

  \begin{figure*}[!tb]
        \centering
        \begin{subfigure}[b]{0.20\textwidth}
            \centering
            \includegraphics[width=0.90\textwidth]{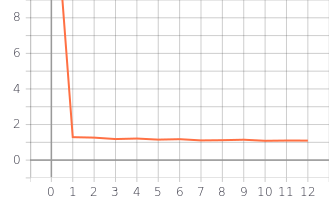}
             \caption{Training set}
            \label{fig:run_train}
        \end{subfigure}
        \begin{subfigure}[b]{0.20\textwidth}  
            \centering 
            \includegraphics[width=0.90\textwidth]{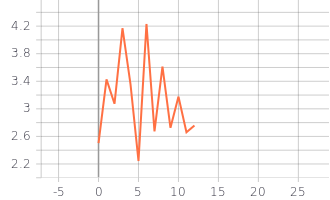}
  \caption{Validation set \#1}
  \label{fig:run_val1}
        \end{subfigure}
        \begin{subfigure}[b]{0.20\textwidth}   
            \centering 
            \includegraphics[width=0.90\textwidth]{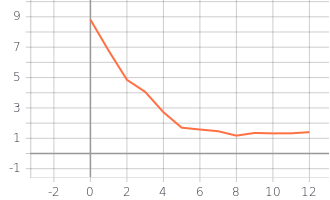}
          \caption{Validation set \#2}
        \label{fig:run_val2}
        \end{subfigure}
        \begin{subfigure}[b]{0.20\textwidth}   
            \centering 
            \includegraphics[width=0.90\textwidth]{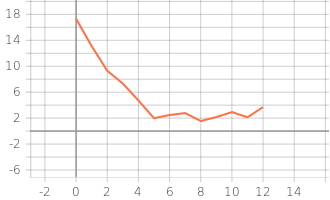}
      \caption{Validation set \#3}
    \label{fig:run_val3}   
        \end{subfigure}
\caption{Model 2 tensorboard run for the training set and validation subsets. The horizontal axis represents the number of epochs, with the vertical axis corresponding to the mean absolute percent error averaged on that epoch. The submission used for the final prediction was the one for epoch \#8. Note that the validation data is the same for each step, while a random 10\% of the training set is evaluated during an epoch.}\label{fig:run_val}
\end{figure*}
\subsection{Extracting Queueing Theory features}
Perhaps the most important aspect of our model is its use of an analytical baseline that serves as a feature extraction step. This algorithm (Algorithm \ref{alg:baseline}) is based on \textit{Queueing Theory} (QT), and iteratively calculates the traffic on links and blocking probabilities. After these iterations, we calculate the \text{traffic intensity} $\rho$, probability of being in state zero $\pi_0$, and {predicted average occupancy} $\bm{L}$. The first two are used as features only in the second model. In addition, we also extract the baseline's per-path-delay prediction, using Equation \ref{eqn:path_delay}.

\ITUpar Our work closely follows the M/M/1/B model used in \cite{rusek2020routenet}. For convenience, we list the same formulas used to model the network. Let $\lambda_{k,i}$ be the amount of traffic from some path $p_k$ passing through
 some link $l_i$. Each path is a sequence of links, and so we can use the notation $\lambda_{k,k(j)}$ to indicate the traffic of path $p_k$ going through the $j$-th link encountered while traversing it. The equations governing the network are:
\begin{align}
      \lambda_{k,i} &= 0, &\text{if $l_i \not\in p_k$}\\
    \lambda_{k,k(1)} &= A_k\\
     \lambda_{k,k(j)} &= A_k \prod_{i=1}^{j-1}(1 - \texttt{Pb}_{k(i)}) &\text{if $j > 1$}\\
     \texttt{Pb}_i &= \frac{(1-\rho_i)\rho_i^{b_i}}{1-\rho_i^{b_i+1}}\\
     \rho_i &= \frac{\sum_{k}\lambda_{k,i}}{c_i}
\end{align}
where: $A_k$ is the demand on the path $p_k$; $b_i$ is the buffer size on link $l_i$; $\texttt{Pb}_i$ is the blocking probability on link $l_i$; $\rho_i$ is the utilization of the link, i.e. the ratio between the total traffic on link $l_i$ and its capacity $c_i$. The dataset of this competition provides us with $b_i$, $c_i$, and $A_k$. Specifically, we have $\forall i: b_i = B = 32$, i.e. all queues can hold up to 32 packets. On the other hand, we must use a fixed point algorithm to iteratively update our estimates of the traffic $\lambda_{k,i}$ and blocking probabilities $\texttt{PB}_i$.

\ITUpar Once all the previous QT quantities have been estimated, we can compute the probability that there isn't a packet in the link's queue:
\begin{equation}
    (\pi_0)_i = \frac{(1-\rho_i)}{1-\rho_i^{b_i+1}}  
\end{equation}
From there, we calculate the average utilization of the link:
\begin{equation}
    \texttt{avg\_utilization}_i = \frac{1}{B}\sum_{j=0}^{B} j(\pi_{0}\rho_i^j)
\end{equation}
where B is the maximum number of packets per queue. Next, the mean packet size is obtained as 
\begin{equation}
    \frac{\texttt{queue\_size}}{B}
\end{equation}
where \texttt{queue\_size} is given as 32000 for this dataset.
 This means that the total delay (waiting and passing through the link) is:
\begin{equation}
    \frac{(x + \sum_{j=0}^{B} j(\pi_{0}\rho_i^j) )}{B} \times \frac{\texttt{queue\_size}}{c_i} 
\end{equation}
with $x=1$. After some experimentation, we found that substituting  $x = \pi_0$ gave slightly better results for this dataset.

\subsection{Hyper-parameters}
The other significant difference between the two models lies in the model size. When developing the second model, we opted to scale down as much as possible. This can be observed by looking at the size (i.e. number of columns) of $\bm{X}_{Lh},\bm{X}_{Ph},\bm{X}_{Nh}$, as well as the hidden layer size for the second multilayer perceptron. The entities that represented individual nodes were entirely discarded, justified partly due to the absence of meaningful attributes for nodes.

\begin{table}[!ht]
\caption{Differences between the two models used. Model 2 uses less hidden input columns, and opts for a simple Linear layer instead of an MLP before message passing.}
\scriptsize
\centering
\begin{tabularx}{0.45\textwidth}{l|X|X|X|}
                 {\textbf{Model}} & \textbf{\# of \newline hidden  \newline input columns} & \textbf{MLP\_1 \newline}                                                                                                    & \textbf{MLP\_2}                                                         \\ \hline
\textbf{1} & $\bm{X}_{Ph}$:64\newline $\bm{X}_{Lh}$:64\newline $\bm{X}_{Nh}$:64                   & Linear(128)\newline LeakyRELU()\newline Linear($3 \times 64$)\newline LeakyRELU()                                               & Linear(512)\newline LeakyRELU()\newline Linear(512)\newline LeakyRELU()\newline Linear(1) \\
\hline 
\textbf{2} & $\bm{X}_{Ph}$:8\newline $\bm{X}_{Lh}$:8\newline                           &
Linear($2 \times 8$)& Linear(128)\newline LeakyRELU()\newline Linear(32)\newline LeakyRELU()\newline Linear(1)
\end{tabularx}
\end{table}
The number of message passing rounds for both models is three. Model 1 uses five baseline iterations, whilst Model 2 reduces that to three iterations. Both models perform message passing iterations for three rounds.
\ITUpar Lastly, we set the initial guess of blocking probabilities to 0.3 when training our models, even though it is most common to set it to zero. This did not seem to affect the convergence of the method. 

\section{TRAINING AND Results}
 We use the Adam optimizer with learning rate equal to \texttt{1e-03}. The batch size is set to 16 . To perform early stopping, we evaluate, after each epoch, on a small subset of each of the three validation sets. Each epoch corresponds to going through some random 10\% of the training set; in addition to this, we select a constant subset of each validation set, sacrificing some accuracy in order to speed up the process.
\ITUpar After each epoch, validation stats are printed and a new model file is saved to the \texttt{./model} folder. We submitted a few models from different epochs. In particular, it seemed that Validation set 1 overestimated the MAPE metric on the test set, whereas validation sets 2 and 3 followed the test set's MAPE more closely. Submissions that prioritized losses on validation sets 2 and 3 were usually more successful (unless the MAPE on validation 1 was significantly large).
\ITUpar Training on Model 1 took just over 8 hours. Training on Model 2 takes just over an hour. The respective model weights were saved. While compiling the initial report, we loaded the model weights and confirmed that they indeed produce the same submissions sent to the challenge.

\ITUpar Tensorboard support was added to the code a few days after training Model 1. It provides another way to look at the performance metrics on-the-fly. The training curves for Model 2 are shown in Fig. \ref{fig:run_val}.

\ITUpar The obtained results are listed in Table \ref{tbl:results}, listing the mean absolute percentage error for the Validation subsets 1/2/3, as well as for the final test set.
\begin{table}[!ht]
\scriptsize

\caption{MAPE error for the validation and test sets. We report results for our 2 model architectures, and their average. In addition, we investigated a few variations of Model 1: using the baseline prediction, not using the higher-level QT features, and not using those features or dividing the original features by \texttt{p\_AvgPktsLambda}}.
\begin{tabularx}{.45\textwidth}{l|X|X|X|X}
                              & \textbf{Val. 1}& \textbf{Val. 2}& \textbf{Val. 3}& \textbf{Test} \\ \hline
\textbf{Model 1}   & 2.71            & 1.33            & 1.65            & 1.45          \\
\textbf{Model 2}   & 3.61            & 1.17            & 1.55            & 1.45          \\
\textbf{Average of predictions} & ---             & ---             & ---             & 1.27          \\ \hline
\textbf{Baseline}             & 12.10           & 9.18            & 9.51            & 10.42             \\
\textbf{M1 w/o QT} & 6.02             & 9.78             & 9.30        &  7.18\\
\textbf{M1 w/o QT or div.} & 8.20            & 45.64             & 250.34        &  85.56
\end{tabularx}
\label{tbl:results}
\end{table}
\ITUpar The final versions of both models performed almost identically on the test set. On the validation sets, Model 1 was better than Model 2 on Validation set 1, and worse on validation sets 2 and 3. Their average was able to attain the lowest MAPE of $1.27$.
\ITUpar A few other models were  evaluated for comparison. The analytical baseline on its own was able to achieve a respectable MAPE of 10.42 on the test set. This is slightly worse than the 7.18 MAPE of Model 1 without the QT features (\texttt{M1 w/o QT}). If we also forget to divide features by \texttt{p\_AvgPktsLambda} as mentioned previously (\texttt{M1 w/o QT or div.}), the model completely fails to generalize to the test set and validation sets 2 and 3. One thing of note is that the training curves can sometimes be unstable: for example, the curve for \texttt{M1 w/o QT} eventually rose to 100 MAPE on Validation set 3. Therefore, having a validation set (even if it consists of just a few examples) is immensely useful to check for generalization and perform early stopping.
\ITUpar
When we don't put any measures in place to generalize to larger graphs, Validation set 1 seems least affected. The bad result of 85.56 MAPE is still better than the 300 reported by the organizer's result for the original Routenet \cite{rnError}. One possibility is that summing the  predictions at each link as a final step leads to better generalization than directly predicting delay on each path.

\ITUpar
The rules of the GNNet Challenge allowed up to 20 submissions. For each submission, the error on the test set was made immediately available to the competitors. Even though we used less than 20 submissions, it is possible that this measure is (slightly) optimistic. Nonetheless, it is apparent that the introduction of higher-level features makes a huge difference when generalizing to larger graphs, and that QT-Routenet outperforms both  Routenet and analytical approaches in this scenario.
\section{Conclusion}
This paper presented a novel approach to generalize per-path-delay predictions to  larger 5G networks. We managed to avoid some of the limitations of graph neural networks by working directly at the input level. Namely, we used a robust baseline based on queueing theory to extract higher-level features, and then fine-tuned them to improve the baseline without sacrificing generalization. We achieved good results in the test set and all 3 validation subsets that exploited different path lengths and link capacities. The proposed solution achieved first place in the GNNet Challenge 2021 and second place overall in the ITU AI/ML in the 5G challenge. We hope that this work will help develop more approaches that fine-tune a robust approximate model that generalizes well to different distributions. For future work, we expect better results when combining our approach with other solutions developed during this challenge. 
\section*{Acknowledgement}
\label{sec:ackn}
\ITUpar This study was financed in part by the Coordenação de Aperfeiçoamento de Nível Superior - Brasil (CAPES) - Finance Code 001.

%%= bibliography =%%
\printbibliography 
%%======%%
\newpage
\ITUpar
\section*{Authors}
\label{sec:auth}

\begin{wrapfigure}{l}{0.32\columnwidth} 
    \vspace{-.1in}
    \includegraphics[width=0.39\columnwidth]{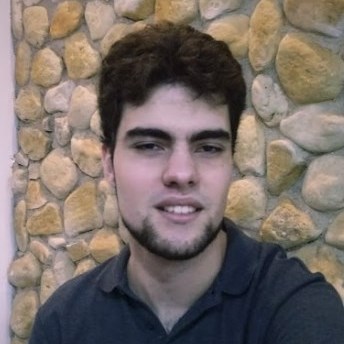} 
\end{wrapfigure}\textbf{Bruno Klaus de Aquino Afonso} is a PhD candidate at the Federal University of São Paulo (ICT-UNIFESP), where he previously received his M.S. and B.S. degrees in computer science. Has been the recipient of fellowships from the Coordination for the Improvement of Higher Education Personnel (CAPES) and The São Paulo Research Foundation (FAPESP). He is interested in making the most out of labels in graph-based semi-supervised learning. While researching GNNs, he stumbled upon ITU’s challenge and the unfamiliar territory of queueing theory and traffic flow in 5G networks. With some perseverance and good fortune, he ended up as the Silver Champion of said challenge.  

\ITUpar

\begin{wrapfigure}{l}{0.32\columnwidth} 
    \vspace{-.1in}
    \includegraphics[width=0.39\columnwidth]{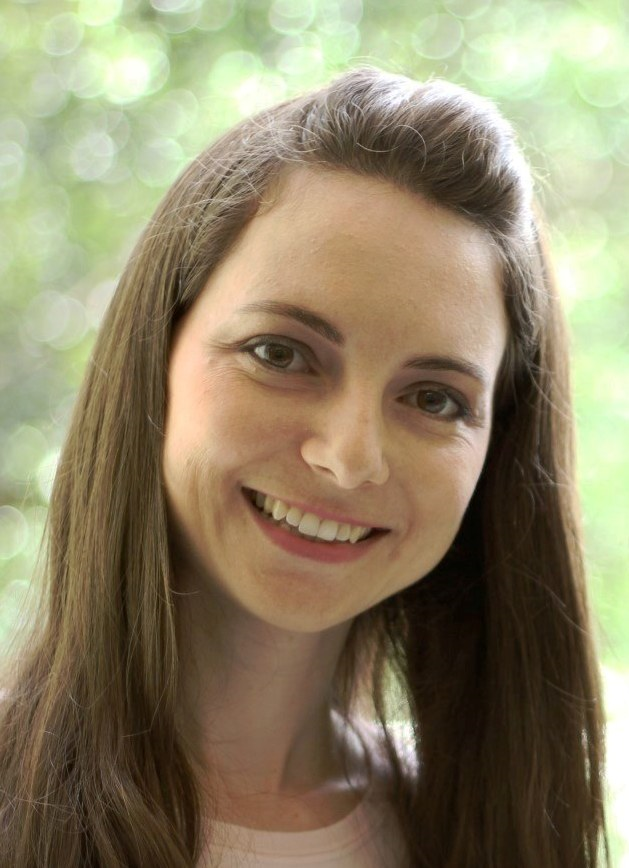} 
\end{wrapfigure}\textbf{Lilian Berton}
 received a B.S. degree in computer science and a licentiate degree in mathematics from Midwest University, Brazil in 2007 and 2008, respectively. She received her M.S. and Ph.D. in computer science from the University of S\~ao Paulo in 2009 and 2016, respectively. She had fellowships from FAPESP. Currently, she is an associate professor at the Institute o Science and Technology - Federal University of S\~ao Paulo, Brazil. Her research interests are machine learning, graph-based methods, complex networks and data mining. \ITUpar
 
\newpage
\section*{Appendix A - Model 1 code}

\begin{algorithm}[H]
\small
\caption{Model 1 (submitted on September 22nd)}\label{alg:cap}
\begin{algorithmic}
\Require{\begin{flalign*}\bm{X} = \texttt{Concatenate}&([\bm{X}_P,\bm{X}_{Ph},\bm{X}_L,\bm{X}_{Lh},\bm{X}_{N},\bm{X}_{Nh}],\texttt{axis=1})\end{flalign*}}
\Require \texttt{B\_path}, \texttt{B\_link}: baseline predictions
\Require $\bm{E}$: network topology 
\Require  \texttt{NUM\_iterations}: \# of message-passing iterations
\algrule
\State $\texttt{E\_lp\_list} \gets \texttt{SeparateEdgeTimeSteps}(\bm{E}_{LP})$
\State $\bm{X} \gets \texttt{MLP\!\_1}(\bm{X},\bm{E}_{LN})$
\For{$0 \leq i < \texttt{NUM\_ITERATIONS}$}
\\ \Comment{Paths receive messages}
\State $\bm{X}_{Ph} \gets \texttt{LeakyRELU}(\texttt{Conv}_{i,\texttt{node\_to\_path}}(\bm{X},\bm{E}_{NP}))$

\State $H \gets \texttt{None}$

\For{$0 \leq k < \texttt{E\_lp\_list} .length$}
\State $H \gets (\texttt{GConvGRU}_{0,\texttt{link\_to\_path}}(\bm{X},H,\texttt{E\_lp\_list[k]}))$
\EndFor
\State $\bm{X}_{Ph} \gets \texttt{LeakyRELU(H/(\texttt{E\_lp\_list}.length))}$
%%%%

\State $(\bm{X}_{Ph})[:,0\!:\!\texttt{B\_path.shape[1]}] \gets \texttt{B\_path}$
%%%%%%
\\ \Comment{Nodes receive messages}
\State $\bm{X}_{Nh} \gets \texttt{LeakyRELU}(\texttt{Conv}_{i,\texttt{path\_to\_node}}(\bm{X},\bm{E}_{PN}))$
\State $\bm{X}_{Nh} \gets \bm{X}_{Nh} + \texttt{LeakyRELU}(\texttt{Conv}_{i,\texttt{link\_to\_node}}(\bm{X},\bm{E}_{LN}))$
\\ \Comment{Links receive messages}
\State $\bm{X}_{Lh} \gets \texttt{LeakyRELU}(\texttt{Conv}_{i,\texttt{node\_to\_link}}(\bm{X},\bm{E}_{NL}))$

\State $\bm{X}_{Lh} \gets \texttt{LeakyRELU}(\texttt{Conv}_{i,\texttt{path\_to\_link}}(\bm{X},\bm{E}_{PL}))$

\State $(\bm{X}_{Lh})\texttt{[:,0:B\_link.shape[1]]} \gets \texttt{B\_link}$

\EndFor
\State $\bm{L} \gets \texttt{Concatenate}(X_L,X_{Lh})$
\State $\bm{L} \gets \texttt{Sigmoid}(\texttt{MLP\_2}(\bm{L}))$\Comment{Predicts average queue utilization}
\\ \Return $\texttt{GetPathDelay}(\bm{L},\bm{E}_{LP})$ \Comment{Obtains per-path-delay}
\end{algorithmic}
\label{alg:model_1}
\end{algorithm}

\section*{Appendix B - Model 2 code}

\begin{algorithm}[H]
\caption{Model 2 (submitted on September 29th)}
\begin{algorithmic}
\small
\Require{$\bm{X} = \texttt{Concatenate([$\bm{X}_P,\!\bm{X}_{Ph},\!\bm{X}_L,\!\bm{X}_{Lh}$]\!,axis=1})$}
\Require \texttt{B\_path}, \texttt{B\_link}: baseline predictions
\Require $\bm{E}$: network topology 
\Require  \texttt{NUM\_iterations}: \# of message-passing iterations
\State $\texttt{E\_lp\_list} \gets \texttt{SeparateEdgeTimeSteps}(\bm{E}_{LP})$
\State $\bm{X} \gets \texttt{MLP\!\_1}(\bm{X},\bm{E}_{LN})$
\For{$0 \leq i < \texttt{NUM\_ITERATIONS}$}
\\ \Comment{Paths receive messages}

\State $H \gets \texttt{None}$

\For{$0 \leq k < \texttt{E\_lp\_list} .length$}
\State $H \gets (\texttt{GConvGRU}_{0,\texttt{link\_to\_path}}(\bm{X},H,\texttt{E\_lp\_list[k]}))$
\EndFor
\State $\bm{X}_{Ph} \gets \texttt{LeakyRELU(H/(\texttt{E\_lp\_list}.length))}$
%%%%

\State $(\bm{X}_{Ph})[:,0\!:\!\texttt{B\_path.shape[1]}] \gets \texttt{B\_path}$
%%%%%%
\\\Comment{Links receive messages}

\State $\bm{X}_{Lh} \gets \texttt{LeakyRELU}(\texttt{Conv}_{i,\texttt{path\_to\_link}}(\bm{X},\bm{E}_{PL}))$

\State $(\bm{X}_{Lh})[:,0\!:\!\texttt{B\_link.shape[1]}] \gets \texttt{B\_link}$

\EndFor
\State $\bm{L} \gets \texttt{Concatenate}(X_L,X_{Lh})$
\State $\bm{L} \gets \texttt{Sigmoid}(\texttt{MLP\_2}(\bm{L}))$\Comment{Predicts average queue utilization}
\\ \Return $\texttt{GetPathDelay}(\bm{L},\bm{E}_{LP})$ \Comment{Obtains per-path-delay}
\end{algorithmic}
\label{alg:model_2}
\end{algorithm}

\section*{Appendix C - Baseline code}

\begin{algorithm}[H]
\small
\caption{Baseline}
\begin{algorithmic}
\Require $\bm{E}$: network topology 
\Require \texttt{p\_PktsGen}: Packets
generated per time unit for each path
\Require \texttt{l\_LinkCapacity}: Vector w/ capacity of each link
\Require \texttt{NUM\_iterations}: number of iterations
\State Initialize $PB$ as a vector with constant values for each link.
\State $B \gets 32$
\State $\texttt{queue\_size} \gets 32000$
\State Let $\lambda_{k,i}$ be the amount of traffic from path k passing through
link i. 
\State $\lambda_{k,k(i)}$ is the traffic from path k passing through its i-th edge.

\For{$0 \leq it < \texttt{NUM\_iterations}$}
\State $A \gets \texttt{p\_PktsGen}$\Comment{$A$ is the demand on each path}
\For{each path $k$}
\State Let $m_k$ be the max number of edges in path $k$.
\State $\lambda_{k,k(1)} \gets A_k$ 
\State $\forall j \in \{1..\leq m_{k}\}: \lambda_{k,k(j)} \gets A_k\prod_{i=1}^{j-1}PB_{i}$
\EndFor
\For{each link $l$}
\Comment{Get total traffic on links}
\State $T_l\gets \sum_{\exists i: l = k(i)} \lambda_{k,l}$
\State $\rho_l \gets T_l / \texttt{l\_LinkCapacity}_l$

\State $\texttt{PB}_l \gets \frac{(1-\rho_l)\rho_l^B} { (1-\rho_l)^{B+1}}$\Comment{Update blocking probabilities}
\EndFor
\EndFor

\State $\pi_0 \gets (1 - \rho)/(1 - \texttt{pow}(\rho,{B+1}))$ \Comment{Prob. that the queue is at state 0}
\State $\bm{L} \gets \frac{1}{B}(\pi_0 + \sum_{j=1}^{B} j (\pi_0 \cdot   \texttt{pow}(\rho,j)))$
\State $\texttt{baseline\_link} \gets [\pi_0,\rho,\bm{L}]$ \Comment{\small Obs: Only $[\bm{L}]$ for Model 1}
\State $\bm{L} \gets \frac{\bm{L}\times \texttt{queue\_size}}{\texttt{l\_LinkCapacity}}$
\State $\texttt{baseline\_path} \gets \texttt{GetPathDelay}(\bm{L},\bm{E}_{LP})$
\State \Return \texttt{baseline\_link}, \texttt{baseline\_path}
\end{algorithmic}
\label{alg:baseline}
\end{algorithm}

\end{document}